\pdfoutput=1

\PassOptionsToPackage{table}{xcolor}
\documentclass[11pt]{article}

\usepackage[final]{acl}

\usepackage{times}
\usepackage{latexsym}

\usepackage[table]{xcolor} 
\usepackage{pdfpages}

\usepackage{algorithm}
\usepackage{algorithmic}

\usepackage[T1]{fontenc}

\usepackage[utf8]{inputenc}

\usepackage{microtype}

\usepackage{inconsolata}

\usepackage{graphicx}

\usepackage{enumitem}

\usepackage{microtype}
\usepackage{booktabs} 

\usepackage{subcaption} 
\usepackage{tikz}
\usepackage{pgf-pie} 
\usepackage{pgfplots}
\usepackage{listings}
\usepackage{tcolorbox}
\usepackage{enumitem}

\definecolor{codegreen}{rgb}{0,0.6,0}
\definecolor{codegray}{rgb}{0.5,0.5,0.5}

\definecolor{backcolour}{RGB}{245,248,250}
\definecolor{emph}{RGB}{166,88,53}
\definecolor{nightblue}{RGB}{9,49,105}
\definecolor{keywords}{RGB}{207,33,46}
\definecolor{lightpurple}{RGB}{130,81,223}
\newcommand{{\methodname}}{UTBoost}
\newcommand{{\utgenname}}{UTGenerator}

\lstdefinestyle{mystyle}{
    backgroundcolor=\color{backcolour},   
    commentstyle=\color{codegreen},
    keywordstyle=\color{keywords},
    stringstyle=\color{nightblue},
    basicstyle=\fontsize{7}{8}\ttfamily,
    breakatwhitespace=true,         
    breaklines=true,                 
    captionpos=b,                    
    keepspaces=true,                 
    numberstyle=\tiny\color{codegray},
    numbersep=2pt,                  
    showspaces=false,                
    showstringspaces=false,
    showtabs=false,                  
    tabsize=2,
    emphstyle={\color{lightpurple}},
    linewidth=1\columnwidth,
    frame=tb,    
    xrightmargin=0pt,
    xleftmargin=0.23cm,
    numbers=left,
    aboveskip=0.2cm,
    belowskip=0.1cm,
}
\title{UTBoost: Rigorous Evaluation of Coding Agents on SWE-Bench}

\author{%
    \textbf{Boxi Yu}$^{1\dag}$
    \quad \textbf{Yuxuan Zhu}$^2$
    \quad \textbf{Pinjia He}$^{1}$\thanks{Corresponding author.}
    \quad \textbf{Daniel Kang}$^2$
    \vspace{1em}
    \\
    $^1$The Chinese University of Hong Kong, Shenzhen  \\
    $^2$University of Illinois Urbana Champaign \\
    $^{1\dag}$\texttt{boxiyu@link.cuhk.edu.cn}
}

\begin{document}
\maketitle

\begin{abstract}
The advent of Large Language Models (LLMs) has spurred the development of coding agents for real-world code generation.
As a widely used benchmark for evaluating the code generation capabilities of these agents, SWE-Bench uses real-world problems based on GitHub issues and their corresponding pull requests.
However, the manually written test cases included in these pull requests are often insufficient, allowing generated patches to pass the tests without resolving the underlying issue.
To address this challenge, we introduce {\utgenname}, an LLM-driven test case generator that automatically analyzes codebases and dependencies to generate test cases for real-world Python projects.
Building on {\utgenname}, we propose {\methodname}, a comprehensive framework for test case augmentation.
In our evaluation, we identified 36 task instances with insufficient test cases and uncovered 345 erroneous patches incorrectly labeled as passed in the original SWE Bench.
These corrections, impacting 40.9\% of SWE-Bench Lite and 24.4\% of SWE-Bench Verified leaderboard entries, yield 18 and 11 ranking changes, respectively.





\end{abstract}
\section{Introduction}

Advances in large language models (LLMs) have enabled the development of automated coding agents capable of generating code for software engineering tasks.
To evaluate their effectiveness on real-world Python projects, prior work introduced SWE-Bench~\cite{jimenez2024swebench}, a benchmark specifically designed for this purpose.
Each instance in SWE-Bench consists of a repository, an issue description, and a set of manually written test cases to verify whether the issue is resolved.
The task of coding agents is to generate a patch that resolves the issue, as demonstrated by successfully passing all relevant test cases.

However, the manually written test cases in SWE-Bench can be too narrow to comprehensively evaluate the correctness of the patches generated by coding agents~\cite{SWE-Bench-Verified, chen2024evaluating, aleithan2024swe}.
Consequently, erroneous patches generated by agents may be incorrectly considered to resolve the issue, compromising the reliability of SWE-Bench.

To comprehensively evaluate the code generation ability of coding agents on real-world Python projects, we propose a novel LLM-based test case generator, {\utgenname}, which automatically generates test cases.
{\utgenname} operates in two steps.
First, it identifies where new test cases should be added by analyzing the codebase and issue description.
Then, based on the location information, {\utgenname} analyzes package dependencies and generates code as unit test cases.


To verify whether the generated patch functions equivalently to the gold patch on the new test cases, we apply intramorphic testing~\cite{rigger2022intramorphic} to construct a test oracle.
Intramorphic testing is a white-box automated testing technique that establishes a test oracle by comparing the outputs of the original and modified systems using the same input.
Since the gold patch and the generated patch are expected to resolve the issue equivalently, the test oracle ensures that both patches pass the same issue-related test cases.

As a motivating example, the issue description in the instance \texttt{mwaskom\_\_seaborn-3010} requires \texttt{PolyFit}, a function that computes polynomial fits for data, to handle missing data in the inputs \texttt{x} and \texttt{y}.
However, the original test case for this issue, as shown in Listing~\ref{lst:oriTest} (line 3), only considers scenarios where both \texttt{x} and \texttt{y} have missing data.
A comprehensive set of tests should include cases where only one of the inputs, \texttt{x} or \texttt{y}, has missing data.
Our solution adds a test case where only \texttt{x} has missing data to complement the original test cases, as shown in Listing~\ref{lst:newTest} (lines 3--4).
Unlike the gold patch that resolves the issue (Listing~\ref{lst:goldPatch}), the generated patch fails to handle these additional cases but throws an error message, as shown in Listing~\ref{lst:ibmPatch} (lines 4--5).
Thus, while the generated patch passes the original test case, it does not resolve the issue.

However, adding new test cases is not sufficient if these test cases are not properly accounted for in the SWE-Bench evaluation pipeline.
This is because SWE-Bench uses a parser based on regular expressions to extract test cases from the test log, but the original parser fails to parse many test cases due to various defects.
For example, it can not handle test cases that span multiple lines in the test log.
To address these issues, we improved the original SWE-Bench parser by fixing these defects.
With the improved parser, we identified 64 erroneous patches generated by coding agents that were incorrectly labeled as passed in SWE-Bench Lite, and 79 erroneous patches that were similarly mislabeled in SWE-Bench Verified.

Building on {\utgenname} and intramorphic testing, we propose {\methodname}, a framework for augmenting test cases in real-world Python projects.
Given a SWE-Bench instance and a generated patch as input, {\methodname} generates new test cases (e.g., Listing~\ref{lst:newTest}) and flags the instance as suspicious if the gold patch and the generated patch behave differently in the new test cases.
If the generated test cases complement the original ones, they are added to the original test suite.


We applied {\methodname} to SWE-Bench Lite~\cite{SWE-Bench-Lite} and SWE-Bench Verified~\cite{SWE-Bench-Verified}.
Our analysis identified 176 erroneous patches in SWE-Bench Lite and 169 in SWE-Bench Verified that were incorrectly evaluated as passing in the original SWE-Bench.
These corrections resulted in leaderboard updates, with ranking changes for 40.9\% of entries in SWE-Bench Lite and 24.4\% in SWE-Bench Verified.
Notably, in the original SWE-Bench Verified leaderboard, \texttt{Amazon-Q-Developer-Agent} ranked 1st and \texttt{devlo} ranked 2nd; however, both now share the 1st rank in the updated leaderboard.
With the augmented test cases generated by {\utgenname}, we identified that 7.7\% (23/300) of instances in SWE-Bench Lite and 5.2\% (26/500) of instances in SWE-Bench Verified have insufficient test cases.
Using our improved parser, we also identified annotation errors in 54.6\% (164/300) of instances in SWE-Bench Lite and 54.2\% (271/500) of instances in SWE-Bench Verified.

    
\begin{figure}
\begin{lstlisting}[language=Python,breaklines=true, caption = {The original test case in SWE-Bench that only considers the case when there is missing data both in \texttt{x} and \texttt{y} (\texttt{mwaskom\_\_seaborn-3010})}., label={lst:oriTest}, showstringspaces=false,literate={í}{{\'i}}1]
def test_missing_data(self, df):
    groupby = GroupBy(["group"])
    df.iloc[5:10] = np.nan
    res1 = PolyFit()(df[["x", "y"]], groupby, 
        "x", {})
    res2 = PolyFit()(df[["x", "y"]].dropna(), 
        groupby, "x", {})
    assert_frame_equal(res1, res2)
\end{lstlisting}
\end{figure}



\begin{figure}
\begin{lstlisting}[language=Python,breaklines=true, caption = {The augmented test case that considers the case when there is only missing data in \texttt{x} (\texttt{mwaskom\_\_seaborn-3010}).}, label={lst:newTest}, showstringspaces=false,literate={í}{{\'i}}1]
def test_none_values(self):
    df = pd.DataFrame({
        "x": [1, 2, 3, None, 4, 5, 6],
        "y": [1, 4, 9, 16, 25, 36, 49],
        "group": [1, 1, 1, 1, 1, 1, 1]
    })
    groupby = GroupBy(["group"])
    res1 = PolyFit()(df, groupby, "x", {})
    res2 = PolyFit()(df.dropna(), groupby, "x", 
        {})
    assert_frame_equal(res1, res2)
\end{lstlisting}
\end{figure}


\begin{figure}[t!]
\begin{lstlisting}[language=Python,breaklines=true, caption = {The gold patch (\texttt{mwaskom\_\_seaborn-3010}).}, label={lst:goldPatch}, showstringspaces=false,literate={í}{{\'i}}1]
def __call__(self, data, groupby, orient, scales):
    return (groupby.apply(
        data.dropna(subset=["x", "y"]), 
        self._fit_predict))
\end{lstlisting}
\end{figure}

\begin{figure}[t!]
\begin{lstlisting}[language=Python,breaklines=true, caption = {The generated patch by \texttt{IBM SWE-1.0} (\texttt{mwaskom\_\_seaborn-3010}).}, label={lst:ibmPatch}, showstringspaces=false,literate={í}{{\'i}}1]
def _fit_predict(self, data):
    y = data["y"].dropna()
    x = data["x"].dropna()
    if x.shape[0] != y.shape[0]:
        raise ValueError("x and y must have the same number of non-missing values")
    if x.nunique() <= self.order:
        # TODO warn?
        xx = yy = []
\end{lstlisting}
\end{figure}
\section{SWE-Bench}

In this section, we introduce SWE-Bench~\cite{jimenez2024swebench} and its two splits: SWE-Bench Lite and SWE-Bench Verified~\cite{SWE-Bench-Verified}.

\paragraph{SWE-Bench}

SWE-Bench is a benchmark for evaluating the code generation capabilities of coding agents on real-world GitHub projects.
It features 12 popular Python repositories and focuses on generating pull requests to address specific issues by producing code edits represented as patch files.
Each task instance includes a gold patch and a set of unit tests crafted by human developers, which serves as a reference for resolving the issue.

In the SWE-Bench evaluation, an agent generates a patch based on the issue description and the codebase.
Then, SWE-Bench evaluates the generated patch using two types of unit tests: \texttt{PASS\_TO\_PASS} and \texttt{FAIL\_TO\_PASS}.
The benchmark evaluates performance by measuring the percentage of patches that successfully pass both types of tests for each instance.
To extract test results from logs generated in instance-specific virtual environments, the evaluation pipeline uses repository-specific parsers with manually crafted regular expressions, averaging 23 lines of code.

\paragraph{SWE-Bench Lite}
The full SWE-bench test split comprises 2,294 issue-commit pairs across 12 Python repositories.
SWE-Bench Lite is a lite version of SWE-Bench, which is a subset of SWE-Bench with 300 task instances.
These instances focus on evaluating functional bug fixes, ensuring they are more self-contained while maintaining the original diversity across 11 of the 12 repositories.

\paragraph{SWE-Bench Verified}

OpenAI introduced a new version of SWE-Bench~\cite{SWE-Bench-Verified}, named SWE-Bench Verified, to improve the robustness and reliability of the evaluation.
They identified two major problems with the data in SWE-Bench:
\begin{itemize}[leftmargin=*]
    \item \textbf{Unit tests}: The unit tests are sometimes too specific or unrelated to the issue, which potentially causes correct solutions to be rejected.
    \item \textbf{Issue description}: Many samples have an issue description that is under-specified, leading to ambiguity on the problem.
\end{itemize}

To address these two issues, OenAI launched a human annotation campaign with 93 professional software developers to verify each sample of the SWE-bench test set for appropriately scoped unit tests and well-specified issue descriptions.
Finally, they released SWE-Bench Verified, a subset of 500 samples that the human annotators verified to be non-problematic.

\section{Methodology}

\begin{figure}[t!]
    \centering
    \includegraphics[width=0.45\textwidth]{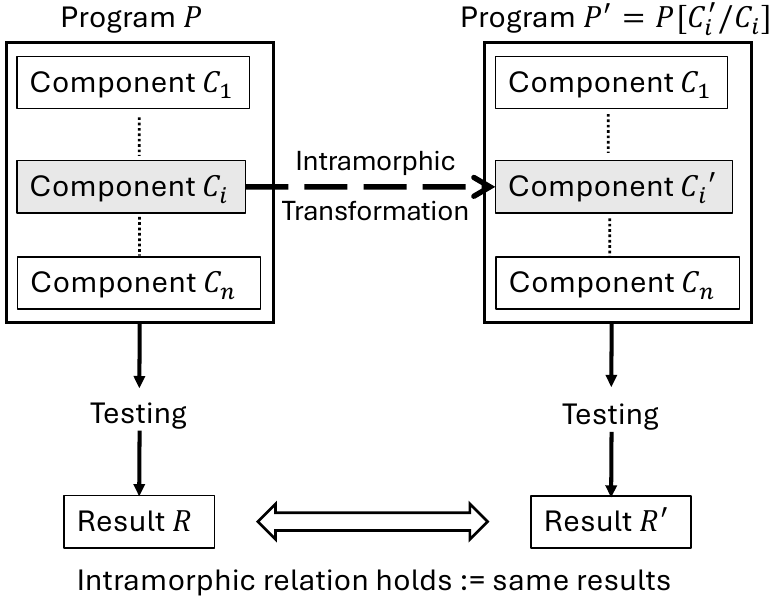}
    \caption{The architecture of intramorphic testing (we define $P$, $C_i$, $R$ as the program, the $i$-th component of the program, and the program's output, respectively).}
    \label{fig:intramorphic}
\end{figure}

\begin{figure*}[t!]
    \centering
    \includegraphics[width=0.95\textwidth]{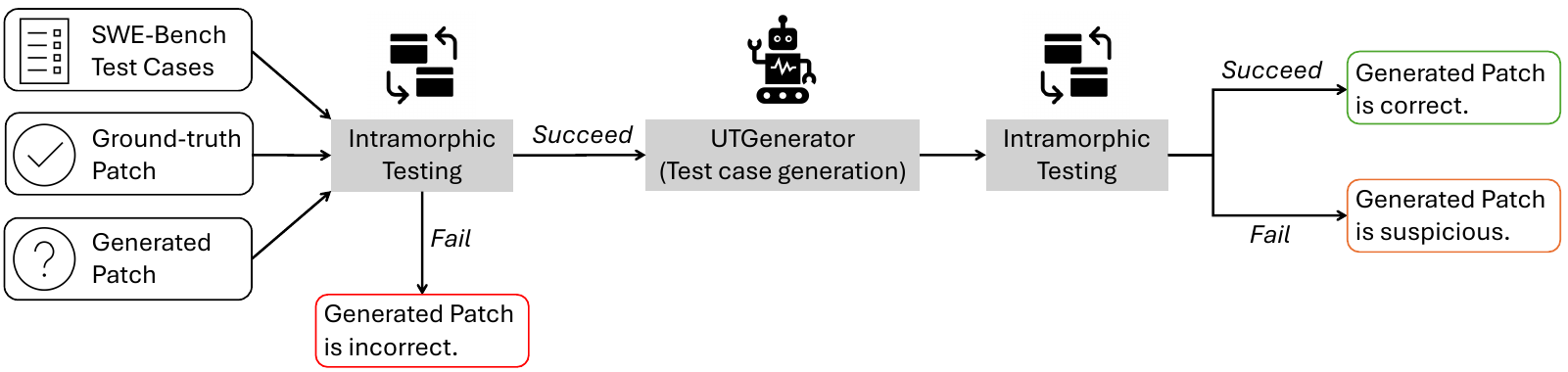}
    \caption{The architecture of {\methodname}.}
    \label{fig:utboost}
\end{figure*}

\begin{figure*}[t!]
    \centering
    \includegraphics[width=0.95\textwidth]{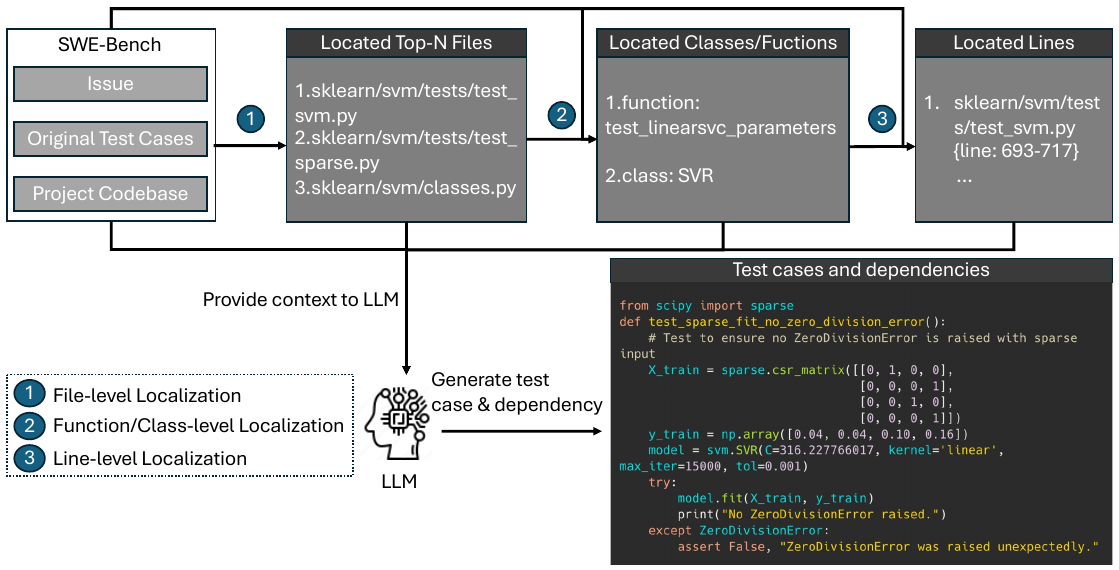}
    \caption{The architecture of {\utgenname}.}
    \label{fig:utgen}
\end{figure*}

In this section, we introduce {\methodname}, a framework for comprehensively testing coding agents using intramorphic testing~\cite{rigger2022intramorphic}.
We discuss the construction of the test oracle, detail the workflow of {\methodname}, and present {\utgenname}, our LLM-based test case generator.

\subsection{Test Oracle}

A test oracle determines whether a system behaves correctly
for a given input.
Automated testing techniques rely on an automated test oracle to test the system without user interaction.
In {\methodname}, we use intramorphic testing to establish a test oracle for evaluating the generated patches.
Intramorphic testing creates a modified version of the system, enabling a single input to define a test oracle that establishes the relationship between the outputs of the original and modified systems.

In SWE-Bench, the gold patch serves as the ground truth for resolving the issue.
A generated patch that resolves the issue provides an alternative implementation to achieve the same functionality as the gold patch and should pass the same test cases associated with the issue.
We define $P$ as the program to which the gold patch is applied and $P'$ as the program to which the generated patch is applied.
The difference between $P$ and $P'$ lies in the component $C$ of $P$, which is transformed into $C'$ in $P'$ through an intramorphic transformation, as illustrated in Figure~\ref{fig:intramorphic}.
We then construct a test oracle to evaluate the generated patches, defined by the intramorphic relation $P(T) = P'(T)$, where $T$ represents the test cases.
To the best of our knowledge, this is the first work to apply intramorphic testing for evaluating real-world software.


\subsection{{\methodname} Workflow}
{\methodname} is an automated testing approach that constructs a test oracle for evaluating generated patches through intramorphic testing, as illustrated in Figure~\ref{fig:utboost}.
We define the original test cases in SWE-Bench as $T_{orig}$ and the augmented test cases as $T_{aug}$.
The {\methodname} process consists of two steps: (1) testing on the original test cases, and (2) testing on the augmented test cases generated by the {\utgenname}.

In the first step, we select the generated patches that pass the original test cases in the same way as gold patches in SWE-Bench, satisfying the intramorphic relation $P(T_{orig}) = P'(T_{orig})$.
We then invoke the test case generator, {\utgenname}, to produce augmented test cases, $T_{aug}$.

In the second step, we apply the augmented test cases $T_{aug}$ to both the program $P$ and the program $P'$ to check whether the intramorphic relation still holds for $T_{aug}$.
If the intramorphic relation $P(T_{aug}) = P'(T_{aug})$ does not hold, we report it as a suspicious issue. This discrepancy indicates that either the gold patch or the generated patch fails to pass the augmented test cases, which implies that the original test cases $T_{orig}$ are insufficient for fully evaluating the patch's correctness.
To achieve a more comprehensive evaluation, we add $T_{aug}$ to the original test suite.




\subsection{\utgenname}

In {\methodname}, a test case generator is required to produce augmented test cases for more comprehensive testing.
To enhance the diversity of these test cases, we introduce {\utgenname}, an LLM-based test case generator.
The architecture of {\utgenname}, illustrated in Figure~\ref{fig:utgen}, consists of two steps: (1) localization and (2) test case generation.
The localization step operates at three levels: file-level, function/class-level, and line-level.


\subsubsection{File-level Localization}

Since real-world project codebases are generally very large, we construct a tree-structured representation of the codebase to organize its files and their locations.
Files and folders at the same directory level are aligned vertically.
{\utgenname} then takes the issue description, the original test patch from SWE-Bench, and the tree-structured codebase as input to an LLM, which identifies the Top-N files most likely to require edits for adding test cases.
Figure~\ref{fig:utgen} illustrates an example of the three levels of file localization in \texttt{scikit-learn\_\_scikit-learn-14894}, an instance from SWE-Bench Verified.
In the first step, {\utgenname} identifies \texttt{sklearn/svm/tests/test\_svm.py} and \texttt{sklearn/svm/base.py} as the most likely files to add the augmented test cases.


\subsubsection{Function/class-level Localization}
For function/class-level localization, we first compress the codebase files by retaining only the headers of classes and functions.
After identifying the Top-N files for potential edits through file-level localization, we provide their compressed formats, along with the issue description and the original test patch, as input to an LLM.
The LLM analyzes these inputs to identify the functions or classes most likely to require augmented test cases.
As illustrated in the second step of Figure~\ref{fig:utgen}, this process identifies the function \texttt{test\_linearsvc\_parameters} and class \texttt{SVR} as the most likely candidates for adding the augmented test cases.


\subsubsection{Line-level Localization}
After identifying the specific functions or classes to add test cases, we extract these code snippets and provide them, along with the issue description and the original test patch, as the input to an LLM.
The LLM analyzes the inputs to determine the specific lines within the functions or classes that are most suitable for adding the augmented test cases.
For instance, as shown in Figure~\ref{fig:utgen}, lines 693--717 of the file \texttt{sklearn/svm/tests/test\_svm.py} are identified as the most likely candidates for adding the augmented test cases.


\subsubsection{Test Case Generation}
The final step is to generate the augmented test cases and their dependencies.
We use a context window of \texttt{x} lines of code to expand the located lines and control the range for adding the augmented test cases.
For example, if the located lines are from line 693 to 717, the context window is defined as \texttt{[max(693-x, 0), min(717+x, end\_line)]}, where \texttt{end\_line} represents the last line of the file.
We then provide the code snippets within this context window, along with the issue description and the original test patch, as the input to an LLM, asking it to generate the augmented test cases and their dependencies.
As shown in Figure~\ref{fig:utgen}, {\utgenname} generates a test case named \texttt{test\_sparse\_fit\_no\_zero\_division\_error} and the corresponding dependency.


\subsection{Improved Parser}

The parser is a critical component of SWE-Bench, responsible for extracting test cases from test logs.
However, the original SWE-Bench parser often failed to parse test cases accurately, particularly when the logs contained side messages or spanned multiple lines.
For instance, in the \texttt{django\_\_django-13710}, the test case \texttt{test\_immutable\_content\_type (admin\_inlines.tests.TestInlineAdminForm)} passes both before and after applying the gold patch.
As shown in Listing~\ref{lst:django_13710_log}, this test case's log spans two lines (lines 2--3).
Due to its limitations, the original parser incorrectly splits the log at the last occurrence of a suffix, erroneously extracting \texttt{"Regression for \#9362"} as the test case name for \texttt{PASS\_TO\_PASS} in \texttt{django\_\_django-13710} (Listing~\ref{lst:swe-parser-django}, line 6).\footnote{\url{https://huggingface.co/datasets/princeton-nlp/SWE-bench\_Lite/}}

To address this issue, we developed an improved parser that robustly handles multi-line test case logs.
Our approach uses a queue to track neighboring log data (Listing~\ref{lst:improved-parser-django}, line 3) and employs regular expressions to accurately match test case names (line 2).
When a test case spans multiple lines, the improved parser iteratively searches until it identifies the correct test case name (Listing~\ref{lst:improved-parser-django}, lines 12--18). For the \texttt{django\_\_django-13710} example, this ensures the correct extraction of the test function \texttt{test\_immutable\_content\_type (admin\_inlines.tests.TestInlineAdminForm)} from the log in Listing~\ref{lst:django_13710_log} (lines 2--3).
Beyond this case, the improved parser addresses multiple limitations of the original SWE-Bench parser, significantly enhancing the reliability and rigor of SWE-Bench.

\begin{figure}[t!]
\begin{lstlisting}[language=,breaklines=true, caption = {Test Log of \texttt{django\_\_django-13710} (before gold patch is applied).}, label={lst:django_13710_log}, showstringspaces=false,literate={í}{{\'i}}1]
# Test log of django__django-13710
test_immutable_content_type (admin_inlines.tests.TestInlineAdminForm)
Regression for #9362 ... ok
test_all_inline_media (admin_inlines.tests.TestInlineMedia) ... ok
\end{lstlisting}
\end{figure}

\begin{figure}[t!]
\begin{lstlisting}[language=Python,breaklines=true, caption = {Original SWE-Bench parser for django.}, label={lst:swe-parser-django}, showstringspaces=false,literate={í}{{\'i}}1]
# SWE-Bench parser for django
pass_suffixes = (" ... ok", " ... OK", " ...  OK")
for suffix in pass_suffixes:
    if line.endswith(suffix):
        ... # omits several lines of code
        test = line.rsplit(suffix, 1)[0]
        test_status_map[test] = TestStatus.PASSED.value
        break
\end{lstlisting}
\end{figure}

\begin{figure}[t!]
\begin{lstlisting}[language=Python,breaklines=true, caption = {Improved parser for Django.}, label={lst:improved-parser-django}, showstringspaces=false,literate={í}{{\'i}}1]
# Improved parser for django
pattern_test = r"[a-zA-Z_]\w*\s\([\w.]+\)"
previous_line = deque()
for line in lines:
    line = line.strip()
    pass_suffixes = (" ... ok", " ... OK", " ...  OK")
    for suffix in pass_suffixes:
        if line.endswith(suffix):
            test = line.rsplit(suffix, 1)[0]
            # process when test log in separate lines
            if not re.fullmatch(pattern_test, test):
                pt = -1
                while previous_line[pt]:
                    if re.fullmatch(pattern_test, previous_line[pt]):
                        test = previous_line[pt]
                        break
                    pt -= 1
            test_status_map[test] = TestStatus.PASSED.value
            break
    previous_line.append(line)
\end{lstlisting}
\end{figure}

\section{Experiments}

In this section, we evaluate the performance of {\methodname} and our improved parser on SWE-Bench. We propose the following three research questions (RQ) to guide our investigation.
\begin{itemize}[leftmargin=*]
    \item RQ1: How effective is {\methodname} in identifying insufficient test cases?
    \item RQ2: How does the parser affect the evaluation of SWE-Bench?
    \item RQ3: How do insufficient test cases and incorrect annotations affect SWE-Bench's leaderboard?
\end{itemize}

\subsection{Experiment Settings}

In our evaluation, we use the generated patches of the coding agents from the official SWE-Bench experiment repository.\footnote{\url{https://github.com/swe-bench/experiments}}
We extract the generated patches of the coding agents that pass the original SWE-Bench tests and evaluate them using our augmented test cases.
Coding agents that do not provide generated patches are excluded from our analysis.
We have released our code and data.\footnote{\url{https://github.com/CUHK-Shenzhen-SE/UTBoost}}

In {\utgenname}, we use GPT-4o (gpt-4o-2024-08-06) as the LLM.
We set a context window of 10 lines of code for test case generation and use Top-3 for file-level localization.
During the localization phase, we use a temperature of 0.8.
In the test case generation phase, we sample one patch with a temperature of 0, 20 patches with a temperature of 0.8, 20 patches with a temperature of 0.9, and 20 patches with a temperature of 0.99.
Lower temperatures (e.g., 0) produce more deterministic and focused outputs, while higher temperatures (e.g., 0.8, 0.9, or 0.99) enable {\utgenname} to generate more diverse and versatile test cases.
In our experiments, using {\utgenname} with temperatures of 0.9 and 0.99 helped to identify additional instances with insufficient test cases, complementing the results obtained with a temperature of 0.8.

When {\methodname} detects discrepancies between the gold patch and generated patches under augmented test cases, two authors with four years of software testing experience manually review the test cases and patches, reaching a consensus on whether the issue stems from inadequate test coverage.
Generating test cases using {\utgenname} costs an average of \$1.6 per SWE-Bench task instance for API usage.
We use cloud servers with Ubuntu 22.04 LTS on CloudLab~\cite{Duplyakin+:ATC19} to evaluate the gold patches and the generated patches on the test cases, which take 300 hours to complete.

\subsection{Effectiveness of {\methodname}}

Overall, we identified 36 task instances with insufficient test cases in SWE-Bench using the augmented test cases generated by {\methodname}.
Of these, 23 instances are from SWE-Bench Lite, and 26 are from SWE-Bench Verified.
We then applied the augmented test cases to evaluate the generated patches that passed the original SWE-Bench test cases.

There are 599 generated patches that pass the 23 instances in SWE-Bench Lite and 584 generated patches that pass the 26 instances in SWE-Bench Verified.
However, our augmented test cases found that 28.4\% (170/599) of the generated patches in SWE-Bench Lite and 15.7\% (92/584) in SWE-Bench Verified are erroneous.

These findings demonstrate that a significant proportion of generated patches recorded as passing in SWE-Bench fail to address the issues effectively because the test cases in SWE-Bench are insufficient. 
This underscores the effectiveness of the augmented test cases generated by {\methodname}.

\begin{figure}[t!]
    \centering
    \begin{subfigure}[t]{0.47\linewidth} 
    \centering
    \begin{tikzpicture}
    \begin{axis}[
        ybar,
        bar width=6pt,
        width=\linewidth,
        height=5cm,
        symbolic x coords={sympy, django, scikit-learn, pylint, matplotlib, seaborn, requests},
        xtick=data,
        xtick style={draw=none},
        x tick label style={rotate=45, anchor=east, font=\scriptsize},
        ylabel={Number of instances},
        ylabel style={font=\scriptsize},
        axis x line*=bottom,
        axis y line*=left,
        ymin=0,
        nodes near coords, 
        nodes near coords align={vertical}, 
        every node near coord/.append style={font=\tiny} 
    ]
    \addplot coordinates {(sympy,8) (django,8) (scikit-learn,2) (pylint,2) (matplotlib,1) (seaborn,1) (requests,1) };
    \end{axis}
    \end{tikzpicture}
    \caption{Insufficient test cases in SWE-Bench Lite}
    \label{fig:lite_insufficient_test_cases}
    \end{subfigure}
    \hfill
    \begin{subfigure}[t]{0.47\linewidth} 
    \centering
    \begin{tikzpicture}
    \begin{axis}[
        ybar,
        bar width=6pt,
        width=\linewidth,
        height=5cm,
        symbolic x coords={sympy, django, scikit-learn, pylint, matplotlib, seaborn, requests},
        xtick=data,
        xtick style={draw=none},
        x tick label style={rotate=45, anchor=east, font=\scriptsize},
        ylabel={Number of patches},
        ylabel style={font=\scriptsize},
        axis x line*=bottom,
        axis y line*=left,
        ymin=0,
        nodes near coords, 
        nodes near coords align={vertical}, 
        every node near coord/.append style={font=\tiny} 
    ]
    \addplot coordinates {(seaborn,7) (sympy,86) (requests,22) (django, 37) (pylint,8) (scikit-learn,5) (matplotlib,5)};
    \end{axis}
    \end{tikzpicture}
    \caption{Erroneous patches in SWE-Bench Lite}
    \label{fig:lite_erroneous_pathces}
    \end{subfigure}

    \begin{subfigure}[t]{0.47\linewidth} 
        \centering
        \begin{tikzpicture}
        \begin{axis}[
            ybar,
            bar width=6pt,
            width=\linewidth,
            height=5cm,
            symbolic x coords={sympy, django, scikit-learn, pylint, matplotlib, sphinx, xarray},
            xtick=data,
            xtick style={draw=none},
            x tick label style={rotate=45, anchor=east, font=\scriptsize},
            ylabel={Number of instances},
            ylabel style={font=\scriptsize},
            axis x line*=bottom,
            axis y line*=left,
            ymin=0,
            nodes near coords, 
            nodes near coords align={vertical}, 
            every node near coord/.append style={font=\tiny} 
        ]
        \addplot coordinates { (django,12) (sympy,6) (scikit-learn,2) (pylint,1) (matplotlib,1) (sphinx, 1) (xarray, 3) };
        \end{axis}
        \end{tikzpicture}
        \caption{Insufficient Test Cases in SWE-Bench Verified}
        \label{fig:verified_insufficient_test_cases}
        \end{subfigure}
        \hfill
        \begin{subfigure}[t]{0.47\linewidth} 
        \centering
        \begin{tikzpicture}
        \begin{axis}[
            ybar,
            bar width=6pt,
            width=\linewidth,
            height=5cm,
            symbolic x coords={sympy, django, scikit-learn, pylint, matplotlib, sphinx, xarray},
            xtick=data,
            xtick style={draw=none},
            x tick label style={rotate=45, anchor=east, font=\scriptsize},
            ylabel={Number of patches},
            ylabel style={font=\scriptsize},
            axis x line*=bottom,
            axis y line*=left,
            ymin=0,
            nodes near coords, 
            nodes near coords align={vertical}, 
            every node near coord/.append style={font=\tiny} 
        ]
        \addplot coordinates { (sympy,46) (django,30) (pylint,1) (scikit-learn,4) (matplotlib,5) (sphinx, 1) (xarray, 5)};
        \end{axis}
        \end{tikzpicture}
        \caption{Erroneous Patches in SWE-Bench Verified}
        \label{fig:verified_erroneous_pathces}
        \end{subfigure}
    
    \caption{Distribution of Insufficient Test Cases And Erroneous Patches.}
    \label{fig:test_cases_distribution}
\end{figure}
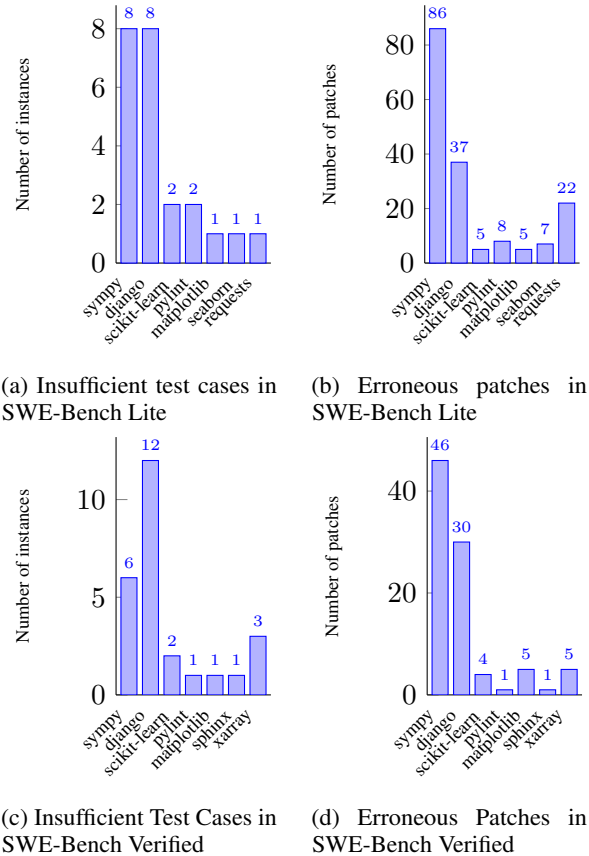

Using {\methodname}, we uncovered insufficient test cases across 9 of the 12 Python projects included in SWE-Bench.
The distribution of the insufficient test cases and erroneous patches are shown in Figure \ref{fig:test_cases_distribution}.
Notably, \texttt{django} and \texttt{sympy} are the most frequent projects with insufficient test cases and erroneous patches in both SWE-Bench Lite and SWE-Bench Verified.
Together, \texttt{django} and \texttt{sympy} account for 84.1\% (143/170) of the erroneous patches in SWE-Bench Lite and 82.6\% (76/92) in SWE-Bench Verified.

\begin{tcolorbox}[boxsep=1pt,left=2pt,right=2pt,top=3pt,bottom=2pt,width=\linewidth,colback=white!90!black,boxrule=0pt, colbacktitle=white!,toptitle=2pt,bottomtitle=1pt,opacitybacktitle=0]
\textbf{Answer to RQ1} \textit{The augmented test cases generated by {\methodname} identified 170 erroneous patches in SWE-Bench Lite and 92 in SWE-Bench Verified that were evaluated as passed by the original test cases, demonstrating the effectiveness of {\methodname} in detecting erroneous patches.}
\end{tcolorbox}

\subsection{Impact of the Parser}

In our evaluation, we found that many annotation errors in SWE-Bench stem from defects in the original SWE-Bench parser.
For instance, 55 test cases in \texttt{django\_\_django-15278}'s \texttt{PASS\_TO\_PASS} are unsuccessfully parsed in SWE-Bench Verified.
We present three selected test cases in the Listing~\ref{lst:bad_parsed} (lines 2--4).
The message \texttt{``Tests altering of the primary key.''} is incorrectly identified as the name of a test case.


\begin{figure}[t!]
\begin{lstlisting}[language=,breaklines=true, caption = { Three selected unsuccessfully parsed test cases in \texttt{django\_\_django-15278} (SWE-Bench Verified).}, label={lst:bad_parsed}, showstringspaces=false,literate={í}{{\'i}}1]
# Three selected unsuccessfully parsed test cases
Tests removing and adding unique_together constraints on a model.
Tries creating a model's table, and then deleting it.
Tests altering of the primary key.
\end{lstlisting}
\end{figure}



\begin{figure}[t!]
    \centering
    
    \begin{subfigure}[t]{0.47\linewidth} 
    \centering
    \begin{tikzpicture}
    \begin{axis}[
        ybar,
        bar width=6pt,
        width=\linewidth,
        height=5cm,
        symbolic x coords={django, sympy, requests, xarray, matplotlib, astropy},
        xtick=data,
        xtick style={draw=none},
        x tick label style={rotate=45, anchor=east, font=\scriptsize},
        ylabel={Number of instances},
        ylabel style={font=\scriptsize},
        axis x line*=bottom,
        axis y line*=left,
        ymin=0,
        nodes near coords, 
        nodes near coords align={vertical}, 
        every node near coord/.append style={font=\tiny} 
    ]
    \addplot coordinates {(sympy,69) (requests,5) (django,84) (xarray,2) (matplotlib,3) (astropy,1)  };
    \end{axis}
    \end{tikzpicture}
    \caption{Erroneous annotations in SWE-Bench Lite}
    \label{fig:lite_fix_parser}
    \end{subfigure}
    \hfill
    \begin{subfigure}[t]{0.47\linewidth} 
    \centering
    \begin{tikzpicture}
    \begin{axis}[
        ybar,
        bar width=6pt,
        width=\linewidth,
        height=5cm,
        symbolic x coords={django, sympy, xarray, matplotlib, requests, astropy, pylint, seaborn},
        xtick=data,
        xtick style={draw=none},
        x tick label style={rotate=45, anchor=east, font=\scriptsize},
        ylabel={Number of patches},
        ylabel style={font=\scriptsize},
        axis x line*=bottom,
        axis y line*=left,
        ymin=0,
        nodes near coords, 
        nodes near coords align={vertical}, 
        every node near coord/.append style={font=\tiny} 
    ]
    \addplot coordinates {(seaborn,1) (sympy,68) (requests,5) (django, 179) (pylint,1) (astropy,2) (matplotlib,7) (xarray,8)};
    \end{axis}
    \end{tikzpicture}
    \caption{Erroneous annotations in SWE-Bench Verified}
    \label{fig:lite_fix_number}
    \end{subfigure}    
    \caption{Distribution of erroneous annotations in SWE-Bench}
    \label{fig:fix_parser}
\end{figure}
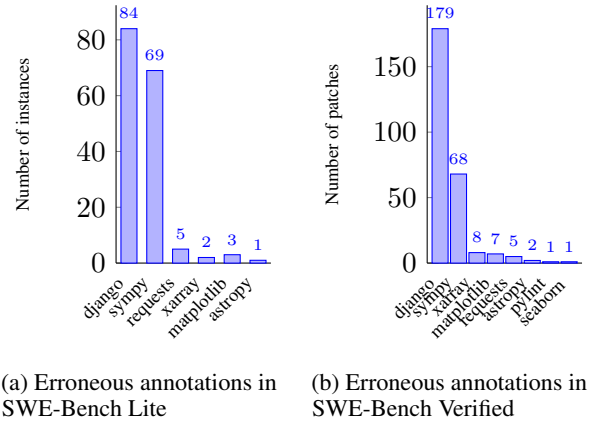

We applied the improved parser to correct the annotation data for \texttt{PASS\_TO\_PASS} and \texttt{FAIL\_TO\_PASS} in SWE-Bench Lite and SWE-Bench Verified.
This update affected 54.7\% (164/300) of the instances in SWE-Bench Lite and 54.2\% (271/500) of the instances in SWE-Bench Verified.
The distribution of instances with erroneous annotations is shown in Figure~\ref{fig:fix_parser}, with \texttt{django} and \texttt{sympy} accounting for the majority of annotation errors.
We evaluated the generated patches using the improved parser with updated annotations, finding that some patches originally marked as passed were actually erroneous.
64 erroneous patches generated by coding agents were incorrectly evaluated as passed in the original SWE-Bench Lite.
Similarly, 79 erroneous patches were incorrectly evaluated as passed in the original SWE-Bench Verified.

\begin{tcolorbox}[boxsep=1pt,left=2pt,right=2pt,top=3pt,bottom=2pt,width=\linewidth,colback=white!90!black,boxrule=0pt, colbacktitle=white!,toptitle=2pt,bottomtitle=1pt,opacitybacktitle=0]
\textbf{Answer to RQ2} \textit{The original parser in SWE-Bench failed to parse many test cases, resulting in incorrect evaluation outcomes.
The improved parser corrected 54.7\% of annotations in SWE-Bench Lite and 54.2\% in SWE-Bench Verified, uncovering 64 erroneous patches in SWE-Bench Lite and 79 in SWE-Bench Verified that were incorrectly classified as passed in the original SWE-Bench evaluation pipeline.}
\end{tcolorbox}

\subsection{Update to the SWE-Bench Leaderboard}



To enhance the accuracy of the SWE-Bench leaderboard, we added the augmented test cases generated by {\methodname} to SWE-Bench and replaced the original SWE-Bench parser with the improved one.
In total, we identified 176 erroneous patches in SWE-Bench Lite and 169 in SWE-Bench Verified that were incorrectly evaluated as passed in the original SWE-Bench.
We recalculated the coding agents' scores on SWE-Bench Lite and SWE-Bench Verified and updated their respective leaderboards accordingly.

With {\methodname}, we found ranking changes in both SWE-Bench Lite and SWE-Bench Verified.
For example, in the original SWE-Bench Verified leaderboard, \texttt{Amazon-Q-Developer-Agent (v20241202-dev)} ranked 1st with a pass@1 rate of 55\%, while \texttt{devlo} ranked 2nd with 54.2\%.
After the update, both agents share the 1st rank with a pass@1 rate of 53.6\%.
This shift occurred because seven patches from \texttt{Amazon-Q-Developer-Agent (v20241202-dev)} were identified as erroneous, compared to only three patches from \texttt{devlo}.
Overall, 40.9\% (18/44) of rankings in SWE-Bench Lite and 24.4\% (11/45) in SWE-Bench Verified changed.
The updated leaderboards for SWE-Bench Lite and SWE-Bench Verified are provided in Appendix~\ref{sec:appendix}.

\begin{tcolorbox}[boxsep=1pt,left=2pt,right=2pt,top=3pt,bottom=2pt,width=\linewidth,colback=white!90!black,boxrule=0pt, colbacktitle=white!,toptitle=2pt,bottomtitle=1pt,opacitybacktitle=0]
    \textbf{Answer to RQ3} \textit{The insufficient test cases and erroneous annotations in SWE-Bench significantly affect the accuracy of the SWE-Bench leaderboard. Updating the leaderboard with augmented test cases and an improved parser results in ranking changes, impacting 40.9\% of SWE-Bench Lite entries and 24.4\% of SWE-Bench Verified entries.}
\end{tcolorbox}

\section{Related Works}

\subsection{Code Generation Benchmark}
Several benchmarks~\cite{chen2021codex, austin2021program} evaluate the code generation ability of LLMs by letting them generate a function or class to solve a problem.
The input for these benchmarks is always straightforward; for example, given an unordered list, the question is to write an algorithm to get the ordered one.
EvalPlus~\cite{liu2024your} adds the augmented test cases via type-aware mutation, such as removing/repeating a random list item.
However, SWE-Bench's test case is more complicated than MBPP and HumanEval because it may involve modification in multiple locations and files, with many dependencies to deal with, e.g., importing functions from other packages.
Therefore, we can not directly apply EvalPlus to add test cases for SWE-Bench since it does not know the locations to add the test cases.
To address the challenges of generating augmented test cases for SWE-Bench, we propose {\utgenname} to consider the codebase and dependencies while generating the test cases.

\subsection{Robustness of SWE-Bench}
The robustness of the coding benchmarks is significant to the rigorous evaluation of the coding agents' ability.
To this end, several works proposed to enhance or discuss the robustness of SWE-Bench.
Aleithan et al.~\cite{aleithan2024swe}, Chen and Jiang~\cite{chen2024evaluating} manually check the passed generated patches of some coding agents and discovered that some of the passed patches are incorrect fixes.
Manually checking these takes lots of time.
Thus, Aleithan et al.~\cite{aleithan2024swe} only check \texttt{SWE-Agent+GPT-4}, and Chen et al.~\cite{chen2024evaluating} select top-10 agents for evaluation.
Comparatively, {\methodname} adds the augmented test cases, which are easy-to-use and applicable for future submissions to SWE-Bench.

OpenAI and SWE-Bench's teams empolyed 93 experienced engineers to manually verify a subset of 500 instances with high quality, which is named as SWE-Bench Verified~\cite{SWE-Bench-Verified}.
However, it is difficult even for experienced engineers to determine if the test case for an issue is comprehensive.
{\methodname} has identified 26 instances with insufficient test cases in SWE-Bench Verified and generated the augmented test cases for them, demonstrating the effectiveness of applying the LLM-based methods to achieve comprehensive testing.
As far as we know, {\methodname} is the first method to address the challenge of insufficient test cases in SWE-Bench, while the existing methods only reveal this problem.
Additionally, we are the first to discuss the impact of parsing errors in the original SWE-Bench evaluatino  harness, which also impedes rigorous evaluation of SWE-Bench.



\section{Conclusion}

In this paper, we introduce {\methodname}, a framework for augmenting test cases in real-world Python projects using intramorphic testing.
Built on {\utgenname}, {\methodname} generates localization- and dependency-aware test cases by analyzing codebases and issue descriptions.
{\methodname} is the first approach to automatically address insufficient test cases in SWE-Bench, identifying 26 instances in SWE-Bench Verified that were overlooked despite a manual review by 93 engineers.

Furthermore, we improved the SWE-Bench parser, uncovering errors in over 54\% of annotations in both SWE-Bench Lite and SWE-Bench Verified.
Using the augmented test cases and improved parser, we identified 176 erroneous patches in SWE-Bench Lite and 169 in SWE-Bench Verified that were incorrectly evaluated as passed in the original SWE-Bench, leading to 40.9\% ranking changes in SWE-Bench Lite and 24.4\% in SWE-Bench Verified.
{\methodname} pioneers the application of intramorphic testing to evaluate open-sourced software systems and provides a versatile framework which has the potential to be adapted for real-world projects in other programming languages.

\section*{Limitations}

{\methodname} augments the test cases of SWE-Bench to achieve robust evaluation.
The main limitation of {\methodname} is that it can only generate test cases for instances that at least one coding agent has resolved because {\methodname} needs to cross-validate the gold patch and generated patches that pass the original SWE-Bench test cases.
Currently, the submitted coding agents have resolved 74.6\% (224/300) and 81.6\% of the test cases in SWE-Bench Lite and SWE-Bench-Verified, for which {\methodname} can generate test cases.

The limitations of our experiments can be summarized in two aspects.  
First, we only used GPT-4o (gpt-4o-2024-08-06) in {\utgenname}. However, integrating other LLM APIs into {\utgenname} is straightforward and could potentially enhance the generation of diverse test cases.  
Second, the architecture of {\utgenname} presents another limitation. Generating test cases and generating patches to address issues are inherently similar tasks, both requiring the identification of relevant locations followed by code generation.
This similarity suggests that adapting a coding agent into a test case generator agent is feasible.
{\utgenname} adopts a simplified architecture inspired by Agentless~\cite{xia2024agentless}, which eliminates the need for the LLM to plan future actions or interact with complex tools.
Currently, there are 48 coding agents and 46 coding agents submitted to the SWE-Bench leaderboard.
Incorporating test case generators with alternative agent frameworks could further diversify the augmented test cases.


\section*{Ethics Statement}
To mitigate the risk of LLMs generating harmful test cases that could compromise software systems, we conduct a thorough manual review of each test case produced by {\methodname}.
This ensures that no harmful code is introduced before integration into the SWE-Bench.
In our paper, we use ChatGPT to check the grammar.
\section*{Acknowledgments}
This paper was supported by the Guangdong Basic and Applied Basic Research Foundation (No. 2024A1515010145), the Shenzhen Science and Technology Program (Shenzhen Key Laboratory Grant No. ZDSYS20230626091302006), and in part by the Open Philanthropy Project.

\bibliography{custom}



\clearpage

\appendix
\section{SWE-Bench Leaderboard}\label{sec:appendix}

In Table~\ref{tab:lite}, we present a comparison between the original and updated leaderboards for SWE-Bench Lite.
Similarly, Table~\ref{tab:verified} shows the comparison for SWE-Bench Verified. Coding agents with ranking changes are highlighted in both tables.
Notably, some coding agents share the same rank, such as \texttt{Amazon Q Developer Agent (v20241202-dev)}~\cite{amazon} and \texttt{devlo}~\cite{devlo}.
Overall, there are 18 ranking changes in SWE-Bench Lite and 11 in SWE-Bench Verified.
The original leaderboards for SWE-Bench Lite and SWE-Bench Verified correspond to the versions dated December 15, 2024.
Since some coding agents do not provide their generated patches, we exclude them, resulting in 44 coding agents in SWE-Bench Lite and 45 in SWE-Bench Verified.

\begin{table*}[t!]
\centering
\resizebox{0.95\textwidth}{!}{

\begin{tabular}{|l|ll|ll|}
\hline
\multicolumn{1}{|c|}{Leadearboard} & \multicolumn{2}{c|}{Original SWE-Bench Lite leaderboard}                                 & \multicolumn{2}{c|}{Updated SWE-Bench Lite leaderboard}                                                                    \\ \hline
Rank                               & \multicolumn{1}{l|}{Coding agent}                                          & \% Resolved & \multicolumn{1}{l|}{Coding agent}                                                          & \% Resolved                   \\ \hline
1                                  & \multicolumn{1}{l|}{Globant Code Fixer Agent~\cite{globant}}                              & 48.33       & \multicolumn{1}{l|}{Globant Code Fixer Agent~\cite{globant}}                                              & \textcolor{red}{46.33}                         \\ \hline
2                                  & \multicolumn{1}{l|}{devlo~\cite{devlo}}                                                 & 47.33       & \multicolumn{1}{l|}{devlo~\cite{devlo}}                                                                 & \textcolor{red}{46.00}                         \\ \hline
3                                  & \multicolumn{1}{l|}{OpenHands + CodeAct v2.1 (claude-3-5-sonnet-20241022)~\cite{codeact_openhand}} & 41.67       & \multicolumn{1}{l|}{OpenHands + CodeAct v2.1 (claude-3-5-sonnet-20241022)~\cite{codeact_openhand}}                 & \textcolor{red}{40.67}                         \\ \hline
4                                  & \multicolumn{1}{l|}{Composio SWE-Kit (2024-10-30)~\cite{composio}}                         & 41.00       & \multicolumn{1}{l|}{Composio SWE-Kit (2024-10-30)~\cite{composio}}                 & \textcolor{red}{39.00}                         \\ \hline
5                                  & \multicolumn{1}{l|}{Agentless-1.5 + Claude-3.5 Sonnet (20241022)~\cite{xia2024agentless}}          & 40.67       & \multicolumn{1}{l|}{Agentless-1.5 + Claude-3.5 Sonnet (20241022)~\cite{xia2024agentless}}                          & \textcolor{red}{38.33}                         \\ \hline
6                                  & \multicolumn{1}{l|}{Bytedance MarsCode Agent~\cite{bytedanceMarscode}}                              & 39.33       & \multicolumn{1}{l|}{Bytedance MarsCode Agent~\cite{bytedanceMarscode}}                                              & \textcolor{red}{38.00}                         \\ \hline
7                                  & \multicolumn{1}{l|}{Moatless Tools + Claude 3.5 Sonnet (20241022)~\cite{moatless}}         & 38.33       & \multicolumn{1}{l|}{Honeycomb~\cite{honycomb}}                                                             & \textcolor{red}{37.67}                         \\ \hline
8                                  & \multicolumn{1}{l|}{Honeycomb~\cite{honycomb}}                                             & 38.33       & \multicolumn{1}{l|}{\cellcolor[HTML]{FFCE93}Moatless Tools + Claude 3.5 Sonnet (20241022)~\cite{moatless}} & \cellcolor[HTML]{FFCE93}\textcolor{red}{36.67} \\ \hline
9                                  & \multicolumn{1}{l|}{AppMap Navie v2~\cite{appmap}}                                       & 36.00       & \multicolumn{1}{l|}{AppMap Navie v2~\cite{appmap}}                                                       & \textcolor{red}{35.00}                         \\ \hline
10                                 & \multicolumn{1}{l|}{Gru(2024-08-11)~\cite{gru}}                                       & 35.67       & \multicolumn{1}{l|}{\cellcolor[HTML]{FFCE93}Isoform~\cite{isoform}}                                       & \cellcolor[HTML]{FFCE93}\textcolor{red}{33.33} \\ \hline
11                                 & \multicolumn{1}{l|}{Isoform~\cite{isoform}}                                               & 35.00       & \multicolumn{1}{l|}{\cellcolor[HTML]{FFCE93}Gru(2024-08-11)~\cite{gru}}                               & \cellcolor[HTML]{FFCE93}\textcolor{red}{33.00} \\ \hline
12                                 & \multicolumn{1}{l|}{SuperCoder2.0~\cite{supercoder}}                                         & 34.00       & \multicolumn{1}{l|}{SuperCoder2.0~\cite{supercoder}}                                                         & \textcolor{red}{32.00}                         \\ \hline
13                                 & \multicolumn{1}{l|}{Alibaba Lingma Agent~\cite{alilingma}}                                  & 33.00       & \multicolumn{1}{l|}{Alibaba Lingma Agent~\cite{alilingma}}                                                  & \textcolor{red}{31.33}                         \\ \hline
14                                 & \multicolumn{1}{l|}{Agentless-1.5 + GPT 4o (2024-05-13)~\cite{xia2024agentless}}                   & 32.00       & \multicolumn{1}{l|}{Agentless-1.5 + GPT 4o (2024-05-13)~\cite{xia2024agentless}}                                   & \textcolor{red}{30.33}                         \\ \hline
15                                 & \multicolumn{1}{l|}{CodeShellTester + GPT 4o (2024-05-13)~\cite{xie2024codeshelltechnicalreport}}                 & 31.33       & \multicolumn{1}{l|}{AutoCodeRover (v20240620) + GPT 4o (2024-05-13)~\cite{zhang2024autocoderover}}                       & \textcolor{red}{30.00}                         \\ \hline
16                                 & \multicolumn{1}{l|}{AutoCodeRover (v20240620) + GPT 4o (2024-05-13)~\cite{zhang2024autocoderover}}       & 30.67       & \multicolumn{1}{l|}{\cellcolor[HTML]{FFCE93}CodeShellTester + GPT 4o (2024-05-13)~\cite{xie2024codeshelltechnicalreport}}         & \cellcolor[HTML]{FFCE93}\textcolor{red}{29.67} \\ \hline
17                                 & \multicolumn{1}{l|}{AIGCode Infant-Coder(2024-08-30)~\cite{aigcode_infant}}                      & 30.00       & \multicolumn{1}{l|}{\cellcolor[HTML]{FFCE93}AIGCode Infant-Coder(2024-08-30)~\cite{aigcode_infant}}              & \cellcolor[HTML]{FFCE93}\textcolor{red}{28.67} \\ \hline
18                                 & \multicolumn{1}{l|}{Amazon Q Developer Agent (v20240719-dev)~\cite{amazon}}              & 29.67       & \multicolumn{1}{l|}{Amazon Q Developer Agent (v20240719-dev)~\cite{amazon}}                              & \textcolor{red}{28.00}                         \\ \hline
19                                 & \multicolumn{1}{l|}{Agentless + RepoGraph + GPT-4o~\cite{xia2024agentless, ouyang2024repograph}}                        & 29.67       & \multicolumn{1}{l|}{Agentless + RepoGraph + GPT-4o~\cite{xia2024agentless, ouyang2024repograph}}                                        & \textcolor{red}{27.67}                         \\ \hline
20                                 & \multicolumn{1}{l|}{CodeR + GPT 4 (1106)~\cite{coder}}                                  & 28.33       & \multicolumn{1}{l|}{CodeR + GPT 4 (1106)~\cite{coder}}                                                  & \textcolor{red}{26.67}                         \\ \hline
21                                 & \multicolumn{1}{l|}{SIMA + GPT 4o (2024-05-13)~\cite{sima}}                            & 27.67       & \multicolumn{1}{l|}{\cellcolor[HTML]{FFCE93}MASAI + GPT 4o (2024-05-13)~\cite{masai}}                   & \cellcolor[HTML]{FFCE93}\textcolor{red}{26.67} \\ \hline
22                                 & \multicolumn{1}{l|}{MASAI + GPT 4o (2024-05-13)~\cite{masai}}                           & 27.33       & \multicolumn{1}{l|}{\cellcolor[HTML]{FFCE93}SIMA + GPT 4o (2024-05-13)~\cite{sima}}                    & \cellcolor[HTML]{FFCE93}\textcolor{red}{26.33} \\ \hline
23                                 & \multicolumn{1}{l|}{Agentless + GPT 4o (2024-05-13)~\cite{xia2024agentless}}                       & 27.33       & \multicolumn{1}{l|}{Agentless + GPT 4o (2024-05-13)}                                       & \textcolor{red}{25.33}                         \\ \hline
24                                 & \multicolumn{1}{l|}{Moatless Tools + Claude 3.5 Sonnet~\cite{moatless}}                    & 26.67       & \multicolumn{1}{l|}{\cellcolor[HTML]{FFCE93}Aider + GPT 4o \& Claude 3 Opus~\cite{aider}}               & \cellcolor[HTML]{FFCE93}\textcolor{red}{25.33} \\ \hline
25                                 & \multicolumn{1}{l|}{OpenHands + CodeAct v1.8~\cite{codeact_openhand}}                              & 26.67       & \multicolumn{1}{l|}{\cellcolor[HTML]{FFCE93}Moatless Tools + Claude 3.5 Sonnet~\cite{moatless}}            & \cellcolor[HTML]{FFCE93}\textcolor{red}{25.33} \\ \hline
26                                 & \multicolumn{1}{l|}{IBM Research Agent-101~\cite{ibm_research_agent}}                                & 26.67       & \multicolumn{1}{l|}{\cellcolor[HTML]{FFCE93}HyperAgent~\cite{hyperagent}}                                    & \cellcolor[HTML]{FFCE93}\textcolor{red}{25.00} \\ \hline
27                                 & \multicolumn{1}{l|}{Aider + GPT 4o \& Claude 3 Opus~\cite{aider}}                       & 26.33       & \multicolumn{1}{l|}{\cellcolor[HTML]{FFCE93}IBM Research Agent-101~\cite{ibm_research_agent}}                        & \cellcolor[HTML]{FFCE93}\textcolor{red}{25.00} \\ \hline
28                                 & \multicolumn{1}{l|}{HyperAgent~\cite{hyperagent}}                                            & 25.33       & \multicolumn{1}{l|}{\cellcolor[HTML]{FFCE93}OpenHands + CodeAct v1.8~\cite{codeact_openhand}}                      & \cellcolor[HTML]{FFCE93}\textcolor{red}{24.33} \\ \hline
29                                 & \multicolumn{1}{l|}{Moatless Tools + GPT 4o (2024-05-13)~\cite{moatless}}                  & 24.67       & \multicolumn{1}{l|}{Moatless Tools + GPT 4o (2024-05-13)~\cite{moatless}}                                  & \textcolor{red}{24.00}                         \\ \hline
30                                 & \multicolumn{1}{l|}{IBM AI Agent SWE-1.0 (with open LLMs)~\cite{ibm_ai_agent}}                  & 23.67       & \multicolumn{1}{l|}{\cellcolor[HTML]{FFCE93}SWE-agent + Claude 3.5 Sonnet~\cite{sweagent}}                 & \cellcolor[HTML]{FFCE93}23.00 \\ \hline
31                                 & \multicolumn{1}{l|}{SWE-agent + Claude 3.5 Sonnet~\cite{sweagent}}                         & 23.00       & \multicolumn{1}{l|}{\cellcolor[HTML]{FFCE93}IBM AI Agent SWE-1.0 (with open LLMs)~\cite{ibm_ai_agent}}         & \cellcolor[HTML]{FFCE93}\textcolor{red}{22.33} \\ \hline
32                                 & \multicolumn{1}{l|}{AppMap Navie + GPT 4o (2024-05-13)~\cite{appmap}}                    & 21.67       & \multicolumn{1}{l|}{AppMap Navie + GPT 4o (2024-05-13)~\cite{appmap}}                                    & \textcolor{red}{20.33}                         \\ \hline
33                                 & \multicolumn{1}{l|}{Bytedance AutoSE (20240828)~\cite{bytedanceMarscode}}                           & 21.67       & \multicolumn{1}{l|}{\cellcolor[HTML]{FFCE93}Bytedance AutoSE (20240828)~\cite{bytedanceMarscode}}                   & \cellcolor[HTML]{FFCE93}\textcolor{red}{19.67} \\ \hline
34                                 & \multicolumn{1}{l|}{Amazon Q Developer Agent (v20240430-dev)~\cite{amazon}}              & 20.33       & \multicolumn{1}{l|}{Amazon Q Developer Agent (v20240430-dev)~\cite{amazon}}                              & \textcolor{red}{19.33}                         \\ \hline
35                                 & \multicolumn{1}{l|}{AutoCodeRover (v20240408) + GPT 4 (0125)~\cite{zhang2024autocoderover}}              & 19.00       & \multicolumn{1}{l|}{AutoCodeRover (v20240408) + GPT 4 (0125)~\cite{zhang2024autocoderover}}                              & \textcolor{red}{18.33}                         \\ \hline
36                                 & \multicolumn{1}{l|}{SWE-agent + GPT 4o (2024-05-13)~\cite{sweagent}}                       & 18.33       & \multicolumn{1}{l|}{\cellcolor[HTML]{FFCE93}SWE-agent + GPT 4 (1106)~\cite{sweagent}}                      & \cellcolor[HTML]{FFCE93}\textcolor{red}{17.33} \\ \hline
37                                 & \multicolumn{1}{l|}{SWE-agent + GPT 4 (1106)~\cite{sweagent}}                              & 18.00       & \multicolumn{1}{l|}{\cellcolor[HTML]{FFCE93}SWE-agent + GPT 4o (2024-05-13)~\cite{sweagent}}               & \cellcolor[HTML]{FFCE93}\textcolor{red}{17.00} \\ \hline
38                                 & \multicolumn{1}{l|}{SWE-agent + Claude 3 Opus~\cite{sweagent}}                             & 11.67       & \multicolumn{1}{l|}{SWE-agent + Claude 3 Opus~\cite{sweagent}}                                             & 11.67                          \\ \hline
39                                 & \multicolumn{1}{l|}{RAG + Claude 3 Opus~\cite{jimenez2024swebench}}                                   & 4.33        & \multicolumn{1}{l|}{RAG + Claude 3 Opus~\cite{jimenez2024swebench}}                                                   & 4.33                          \\ \hline
40                                 & \multicolumn{1}{l|}{RAG + Claude 2~\cite{jimenez2024swebench}}                                        & 3.00        & \multicolumn{1}{l|}{RAG + Claude 2~\cite{jimenez2024swebench}}                                                        & 3.00                          \\ \hline
41                                 & \multicolumn{1}{l|}{RAG + GPT 4 (1106)~\cite{jimenez2024swebench}}                                    & 2.67        & \multicolumn{1}{l|}{RAG + GPT 4 (1106)~\cite{jimenez2024swebench}}                                                    & \textcolor{red}{2.33}                          \\ \hline
42                                 & \multicolumn{1}{l|}{RAG + SWE-Llama 7B~\cite{jimenez2024swebench}}                                    & 1.33        & \multicolumn{1}{l|}{RAG + SWE-Llama 7B~\cite{jimenez2024swebench}}                                                    & 1.00                          \\ \hline
43                                 & \multicolumn{1}{l|}{RAG + SWE-Llama 13B~\cite{jimenez2024swebench}}                                   & 1.00        & \multicolumn{1}{l|}{\cellcolor[HTML]{FFCE93}RAG + SWE-Llama 13B~\cite{jimenez2024swebench}}                           & \cellcolor[HTML]{FFCE93}1.00  \\ \hline
44                                 & \multicolumn{1}{l|}{RAG + ChatGPT 3.5~\cite{jimenez2024swebench}}                                     & 0.33        & \multicolumn{1}{l|}{RAG + ChatGPT 3.5~\cite{jimenez2024swebench}}                                                     & 0.33                          \\ \hline
\end{tabular}

}
\caption{Comparison between the original and updated SWE-Bench Lite leaderboard (We highlight the background for agents with ranking changes and the text for agents whose percentage of resolved cases has changed).} 
\label{tab:lite} 

\end{table*}

\begin{table*}[t!]
\centering
\resizebox{0.95\textwidth}{!}{
\begin{tabular}{|l|ll|ll|}
\hline
\multicolumn{1}{|c|}{Leadearboard} & \multicolumn{2}{c|}{Original SWE-Bench Verifiied leaderboard}                                         & \multicolumn{2}{c|}{Updated SWE-Bench Verified leaderboard}                                                                          \\ \hline
Rank                               & \multicolumn{1}{l|}{Coding agent}                                                       & \% Resolved & \multicolumn{1}{l|}{Coding agent}                                                                    & \% Resolved                   \\ \hline
1                                  & \multicolumn{1}{l|}{Amazon Q Developer Agent (v20241202-dev)~\cite{amazon}}                           & 55.00       & \multicolumn{1}{l|}{Amazon Q Developer Agent (v20241202-dev)~\cite{amazon}}                                        & \textcolor{red}{53.60}                         \\ \hline
2                                  & \multicolumn{1}{l|}{devlo~\cite{devlo}}                                                              & 54.20       & \multicolumn{1}{l|}{\cellcolor[HTML]{FFCE93}devlo~\cite{devlo}}                                                   & \cellcolor[HTML]{FFCE93}\textcolor{red}{53.60} \\ \hline
3                                  & \multicolumn{1}{l|}{OpenHands + CodeAct v2.1 (claude-3-5-sonnet-20241022)~\cite{codeact_openhand}}              & 53.00       & \multicolumn{1}{l|}{OpenHands + CodeAct v2.1 (claude-3-5-sonnet-20241022)~\cite{codeact_openhand}}                           & \textcolor{red}{51.80}                         \\ \hline
4                                  & \multicolumn{1}{l|}{Engine Labs (2024-11-25)~\cite{engine}}                                           & 51.80       & \multicolumn{1}{l|}{Engine Labs (2024-11-25)~\cite{engine}}                                                        & \textcolor{red}{50.80}                         \\ \hline
5                                  & \multicolumn{1}{l|}{Agentless-1.5 + Claude-3.5 Sonnet (20241022)~\cite{xia2024agentless}}                       & 50.80       & \multicolumn{1}{l|}{Agentless-1.5 + Claude-3.5 Sonnet (20241022)~\cite{xia2024agentless}}                                    & \textcolor{red}{49.60}                         \\ \hline
6                                  & \multicolumn{1}{l|}{Solver (2024-10-28)~\cite{solver}}                                                & 50.00       & \multicolumn{1}{l|}{Bytedance MarsCode Agent~\cite{bytedanceMarscode}}                                                        & \textcolor{red}{49.40}                         \\ \hline
7                                  & \multicolumn{1}{l|}{Bytedance MarsCode Agent~\cite{bytedanceMarscode}}                                           & 50.00       & \multicolumn{1}{l|}{\cellcolor[HTML]{FFCE93}Solver (2024-10-28)~\cite{solver}}                                     & \cellcolor[HTML]{FFCE93}\textcolor{red}{49.20} \\ \hline
8                                  & \multicolumn{1}{l|}{nFactorial (2024-11-05)~\cite{nfactorialai}}                                            & 49.20        & \multicolumn{1}{l|}{nFactorial (2024-11-05)~\cite{nfactorialai}}                                                         & \textcolor{red}{48.40}                         \\ \hline
9                                  & \multicolumn{1}{l|}{Tools + Claude 3.5 Sonnet (2024-10-22)~\cite{tools_claude}}                             & 49.00       & \multicolumn{1}{l|}{Tools + Claude 3.5 Sonnet (2024-10-22)~\cite{tools_claude}}                                          & \textcolor{red}{48.20}                         \\ \hline
10                                 & \multicolumn{1}{l|}{Composio SWE-Kit (2024-10-25)~\cite{composio}}                                      & 48.60       & \multicolumn{1}{l|}{Composio SWE-Kit (2024-10-25)~\cite{composio}}                                                   & \textcolor{red}{47.40}                         \\ \hline
11                                 & \multicolumn{1}{l|}{AppMap Navie v2~\cite{appmap}}                                                    & 47.20       & \multicolumn{1}{l|}{AppMap Navie v2~\cite{appmap}}                                                                 & \textcolor{red}{46.40}                         \\ \hline
12                                 & \multicolumn{1}{l|}{Emergent E1 (v2024-10-12)~\cite{emergent}}                                          & 46.60       & \multicolumn{1}{l|}{Emergent E1 (v2024-10-12)~\cite{emergent}}                                                       & \textcolor{red}{45.60}                         \\ \hline
13                                 & \multicolumn{1}{l|}{AutoCodeRover-v2.0 (Claude-3.5-Sonnet-20241022)~\cite{zhang2024autocoderover}}                    & 46.20       & \multicolumn{1}{l|}{AutoCodeRover-v2.0 (Claude-3.5-Sonnet-20241022)~\cite{zhang2024autocoderover}}                                 & \textcolor{red}{45.40}                         \\ \hline
14                                 & \multicolumn{1}{l|}{Solver (2024-09-12)~\cite{solver}}                                                & 45.40       & \multicolumn{1}{l|}{Solver (2024-09-12)~\cite{solver}}                                                             & \textcolor{red}{44.80}                         \\ \hline
15                                 & \multicolumn{1}{l|}{Gru(2024-08-24)~\cite{gru}}                                                    & 45.20       & \multicolumn{1}{l|}{Gru(2024-08-24)~\cite{gru}}                                                                 & \textcolor{red}{43.8}                          \\ \hline
16                                 & \multicolumn{1}{l|}{Solver (2024-09-12)~\cite{solver}}                                                & 43.60       & \multicolumn{1}{l|}{Solver (2024-09-12)~\cite{solver}}                                                             & \textcolor{red}{42.80}                         \\ \hline
17                                 & \multicolumn{1}{l|}{nFactorial (2024-10-30)~\cite{nfactorialai}}                                            & 41.60       & \multicolumn{1}{l|}{nFactorial (2024-10-30)~\cite{nfactorialai}}                                                         & \textcolor{red}{40.60}                         \\ \hline
18                                 & \multicolumn{1}{l|}{Nebius AI Qwen 2.5 72B Generator + LLama 3.1 70B Critic~\cite{nebius}}            & 40.60       & \multicolumn{1}{l|}{Honeycomb~\cite{honycomb}}                                                                       & \textcolor{red}{40.20}                         \\ \hline
19                                 & \multicolumn{1}{l|}{Tools + Claude 3.5 Haiku~\cite{tools_claude}}                                           & 40.60       & \multicolumn{1}{l|}{\cellcolor[HTML]{FFCE93}Tools + Claude 3.5 Haiku~\cite{tools_claude}}                                & \cellcolor[HTML]{FFCE93}\textcolor{red}{40.00} \\ \hline
20                                 & \multicolumn{1}{l|}{Honeycomb~\cite{honycomb}}                                                          & 40.60       & \multicolumn{1}{l|}{\cellcolor[HTML]{FFCE93}Nebius AI Qwen 2.5 72B Generator + LLama 3.1 70B Critic~\cite{nebius}} & \cellcolor[HTML]{FFCE93}\textcolor{red}{39.60} \\ \hline
21                                 & \multicolumn{1}{l|}{Composio SWEkit + Claude 3.5 Sonnet (2024-10-16)~\cite{composio}}                   & 40.60       & \multicolumn{1}{l|}{\cellcolor[HTML]{FFCE93}Composio SWEkit + Claude 3.5 Sonnet (2024-10-16)~\cite{tools_claude}}        & \cellcolor[HTML]{FFCE93}\textcolor{red}{39.00} \\ \hline
22                                 & \multicolumn{1}{l|}{EPAM AI/Run Developer Agent v20241029 + Anthopic Claude 3.5 Sonnet~\cite{epam}} & 39.60       & \multicolumn{1}{l|}{EPAM AI/Run Developer Agent v20241029 + Anthopic Claude 3.5 Sonnet~\cite{epam}}              & \textcolor{red}{39.00}                         \\ \hline
23                                 & \multicolumn{1}{l|}{Amazon Q Developer Agent (v20240719-dev)~\cite{amazon}}                           & 38.80        & \multicolumn{1}{l|}{Agentless-1.5 + GPT 4o (2024-05-13)~\cite{xia2024agentless}}                                             & \textcolor{red}{38.40}                         \\ \hline
24                                 & \multicolumn{1}{l|}{Agentless-1.5 + GPT 4o (2024-05-13)~\cite{xia2024agentless}}                                & 38.80        & \multicolumn{1}{l|}{\cellcolor[HTML]{FFCE93}Amazon Q Developer Agent (v20240719-dev)~\cite{amazon}}                & \cellcolor[HTML]{FFCE93}\textcolor{red}{38.00} \\ \hline
25                                 & \multicolumn{1}{l|}{AutoCodeRover (v20240620) + GPT 4o (2024-05-13)~\cite{zhang2024autocoderover}}                    & 38.40        & \multicolumn{1}{l|}{AutoCodeRover (v20240620) + GPT 4o (2024-05-13)~\cite{zhang2024autocoderover}}                                 & \textcolor{red}{37.80}                         \\ \hline
26                                 & \multicolumn{1}{l|}{Artemis Agent v1 (2024-11-20)~\cite{artemis}}                                      & 32.00       & \multicolumn{1}{l|}{Artemis Agent v1 (2024-11-20)~\cite{artemis}}                                                   & \textcolor{red}{30.80}                         \\ \hline
27                                 & \multicolumn{1}{l|}{nFactorial (2024-10-07)~\cite{nfactorialai}}                                            & 31.60       & \multicolumn{1}{l|}{nFactorial (2024-10-07)~\cite{nfactorialai}}                                                         & \textcolor{red}{30.80}                         \\ \hline
28                                 & \multicolumn{1}{l|}{Lingma Agent + Lingma SWE-GPT 72b (v0925)~\cite{alilingma}}                          & 28.80       & \multicolumn{1}{l|}{Lingma Agent + Lingma SWE-GPT 72b (v0925)~\cite{alilingma}}                                       & \textcolor{red}{27.20}                         \\ \hline
29                                 & \multicolumn{1}{l|}{EPAM AI/Run Developer Agent + GPT4o~\cite{epam}}                                & 27.00       & \multicolumn{1}{l|}{EPAM AI/Run Developer Agent + GPT4o~\cite{epam}}                                             & \textcolor{red}{26.80}                         \\ \hline
30                                 & \multicolumn{1}{l|}{AppMap Navie + GPT 4o (2024-05-13)~\cite{appmap}}                                 & 26.20       & \multicolumn{1}{l|}{\cellcolor[HTML]{FFCE93}nFactorial (2024-10-01)~\cite{nfactorialai}}                                 & \cellcolor[HTML]{FFCE93}\textcolor{red}{25.60} \\ \hline
31                                 & \multicolumn{1}{l|}{nFactorial (2024-10-01)~\cite{nfactorialai}}                                            & 25.80       & \multicolumn{1}{l|}{\cellcolor[HTML]{FFCE93}AppMap Navie + GPT 4o (2024-05-13)~\cite{appmap}}                      & \cellcolor[HTML]{FFCE93}\textcolor{red}{25.20} \\ \hline
32                                 & \multicolumn{1}{l|}{Amazon Q Developer Agent (v20240430-dev)~\cite{amazon}}                           & 25.60       & \multicolumn{1}{l|}{\cellcolor[HTML]{FFCE93}Lingma Agent + Lingma SWE-GPT 72b (v0918)~\cite{alilingma}}               & \cellcolor[HTML]{FFCE93}\textcolor{red}{24.80} \\ \hline
33                                 & \multicolumn{1}{l|}{Lingma Agent + Lingma SWE-GPT 72b (v0918)~\cite{alilingma}}                          & 25.00       & \multicolumn{1}{l|}{Amazon Q Developer Agent (v20240430-dev)~\cite{amazon}}                                        & \textcolor{red}{24.80}                         \\ \hline
34                                 & \multicolumn{1}{l|}{EPAM AI/Run Developer Agent + GPT4o~\cite{epam}}                                & 24.00       & \multicolumn{1}{l|}{EPAM AI/Run Developer Agent + GPT4o~\cite{epam}}                                             & \textcolor{red}{23.80}                         \\ \hline
35                                 & \multicolumn{1}{l|}{SWE-agent + GPT 4o (2024-05-13)~\cite{sweagent}}                                    & 23.20       & \multicolumn{1}{l|}{SWE-agent + GPT 4o (2024-05-13)~\cite{sweagent}}                                                 & \textcolor{red}{22.40}                         \\ \hline
36                                 & \multicolumn{1}{l|}{SWE-agent + GPT 4 (1106)~\cite{sweagent}}                                           & 22.40       & \multicolumn{1}{l|}{SWE-agent + GPT 4 (1106)~\cite{sweagent}}                                                        & \textcolor{red}{21.80}                         \\ \hline
37                                 & \multicolumn{1}{l|}{SWE-agent + Claude 3 Opus~\cite{sweagent}}                                          & 18.20       & \multicolumn{1}{l|}{SWE-agent + Claude 3 Opus~\cite{sweagent}}                                                       & \textcolor{red}{17.80}                         \\ \hline
38                                 & \multicolumn{1}{l|}{Lingma Agent + Lingma SWE-GPT 7b (v0925)~\cite{alilingma}}                           & 18.20       & \multicolumn{1}{l|}{Lingma Agent + Lingma SWE-GPT 7b (v0925)~\cite{alilingma}}                                        & \textcolor{red}{17.80}                         \\ \hline
39                                 & \multicolumn{1}{l|}{Lingma Agent + Lingma SWE-GPT 7b (v0918)~\cite{alilingma}}                           & 10.20       & \multicolumn{1}{l|}{Lingma Agent + Lingma SWE-GPT 7b (v0918)}                                        & \textcolor{red}{9.60}                          \\ \hline
40                                 & \multicolumn{1}{l|}{RAG + Claude 3 Opus~\cite{jimenez2024swebench}}                                                & 7.00        & \multicolumn{1}{l|}{RAG + Claude 3 Opus~\cite{jimenez2024swebench}}                                                             & 7.00                          \\ \hline
41                                 & \multicolumn{1}{l|}{RAG + Claude 2~\cite{jimenez2024swebench}}                                                     & 4.40        & \multicolumn{1}{l|}{RAG + Claude 2~\cite{jimenez2024swebench}}                                                                  & \textcolor{red}{4.20}                          \\ \hline
42                                 & \multicolumn{1}{l|}{RAG + GPT 4 (1106)~\cite{jimenez2024swebench}}                                                 & 2.80        & \multicolumn{1}{l|}{RAG + GPT 4 (1106)~\cite{jimenez2024swebench}}                                                              & \textcolor{red}{2.60}                          \\ \hline
43                                 & \multicolumn{1}{l|}{RAG + SWE-Llama 7B~\cite{jimenez2024swebench}}                                                 & 1.40        & \multicolumn{1}{l|}{\cellcolor[HTML]{FFCE93}RAG + SWE-Llama 13B~\cite{jimenez2024swebench}}                                     & \cellcolor[HTML]{FFCE93}\textcolor{red}{1.00}  \\ \hline
44                                 & \multicolumn{1}{l|}{RAG + SWE-Llama 13B~\cite{jimenez2024swebench}}                                                & 1.20        & \multicolumn{1}{l|}{\cellcolor[HTML]{FFCE93}RAG + SWE-Llama 7B~\cite{jimenez2024swebench}}                                      & \cellcolor[HTML]{FFCE93}\textcolor{red}{0.80}  \\ \hline
45                                   & \multicolumn{1}{l|}{RAG + ChatGPT 3.5~\cite{jimenez2024swebench}}                                                  & 0.40        & \multicolumn{1}{l|}{RAG + ChatGPT 3.5~\cite{jimenez2024swebench}}                                                               & 0.40                          \\ \hline
\end{tabular}

}
\caption{Comparison between the original and updated SWE-Bench Verified leaderboard (We highlight the background for agents with ranking changes and the text for agents whose percentage of resolved cases has changed).} 
\label{tab:verified} 

\end{table*}


\end{document}